\documentstyle[12pt,epsf]{article}

\title{Gravitational effects in terms of paths\\ in Minkowski 
space\thanks{Published in ``Gravity and Cosmology'', vol. 8, 
Suppl. 2, pp. 102-108 (2002)
}}
\author{Michael B. Mensky\\
{\small P.N.Lebedev Physics Institute, 117924 Moscow, Russia}}
\date{}
\newcommand{\Sect}[1]{Sect.~\ref{#1}}
\newcommand{\Fig}[1]{Fig.~\ref{#1}}
\newcommand{\partderiv}[2]{\frac{\partial #1}{\partial #2}}
\newcommand{\al}{\alpha}
\newcommand{\be}{\begin{equation}}
\newcommand{\ee}{\end{equation}}
\newcommand{\ba}{\begin{eqnarray}}
\newcommand{\ea}{\end{eqnarray}}
\newcommand{\bi}{\begin{itemize}}
\newcommand{\ei}{\end{itemize}}

\newcommand{\M}{{\cal M}}
\newcommand{\X}{{\cal X}}
\newcommand{\N}{{\cal N}}
\newcommand{\B}{{\cal B}}

\begin{document}

\maketitle

\begin{abstract}
An approach is developed which enables one to analyze gravitational
effects without usage of any concrete model of geometry (or a class of
models of geometry) of space-time and even without any coordinate
system. Instead, the formalism of the group of paths is used. An
element of this group (a class of curves in Minkowski space) is
associated with each curve in the curved space-time and called its
`flat model'. The analysis of observational data in terms of flat
models of closed curves is interpreted as a formalization of the
analysis which a `naive observer' (knowing nothing about the
space-time being curved) applies to his observations.
\end{abstract}

\section{Introduction}

Gravitational experiments and gravitational effects are usually
described and analyzed in the framework of a certain assumption about
the geometry of space-time, its belonging to a rather narrow class of
geometries \cite{Wheeler95,Lamm00}. The post-Newtonian approximation
for the analysis of gravitational effects in the solar system and
Friedmann model for the analysis of data about early Universe are
examples of this situation. Sometimes the geometrical model used in
the analysis enables one to ascribe a direct physical (geometrical)
sense to a certain coordinate system.

However, such a model-dependent analysis is inconvenient if there is
no sufficient ground for a reasonable choice of a geometrical model of
the space-time region under investigation. In the extreme situation
even topology of this region may be indefinite. Then one cannot
introduce even structure of manifold, to say nothing about the
curvature of the space-time. Even coordinate systems cannot be then
correctly introduced. Such a situation may arise in astrophysics and
cosmology if regions with strong gravitational field or early stages
of Universe are under investigation.

Of course, one may use a number of radically different geometrical
models in this case, trying to analyze the data in the framework of
each of these models. Nevertheless, any a priori choice of the models
may restrict the scope of research, introduce a subjective element
into it. An `objective' approach making use of no concrete model of
geometry may be advantageous in this situation. We shall present here
such an `objective' description of gravitational effects which is
independent of any concrete model of geometry. Particularly, no at all
coordinate system (in a curved space-time) is used in this approach.

At first sight it seems impossible to do without any model of geometry
and even without any coordinate system. However, the following
argument shows that the procedures of this type must exist. Indeed, an
observer can objectively present his observations if he describes in
detail the measuring devices which have been used, his actions with
these devices and the readouts obtained in the course of the
measurements. In other words, a sort of the measurement protocol
presents the observational data in the objective way, with no
reference to models of geometry and coordinate systems.

This way of objectification is of course inconvenient from theoretical
point of view because it makes use of terms and concepts which are too
different from those typical for geometry. We shall present the way of
description of observational data which is adequate for geometry, yet
quite similar to the measurement protocol. This description is based
on the formalism of {\em paths in Minkowski space}. This formalism is
closely connected with one used in the path-dependent field theory
suggested by Mandelstam and Bialynicki-Birula
\cite{Mandelstam,Birula63} and in the path-group approach to gauge
theory and gravity suggested by the author
\cite{FlatModel72,TMF74,PathGr83,PathGr88}.

Although the formalism of paths is mathematically well elaborated, we
shall only shortly concern here the precise mathematical concepts in
\Sect{SectMath}. To avoid unessential mathematical details, we shall
remain in most part of the paper on the `intuitive' level. The
resulting picture is in fact the point of view of a `naive' observer
who attempts to present the events, which are actually performed in
the curved space-time, in terms of his local (and therefore flat)
geometry. We shall call the corresponding procedure `flat modelling'
of a curved space-time. More precisely, the procedure provides flat
models of curves which are given in a curved space-time. It will be
especially important for us to construct flat models for those {\em
closed curves} in a curved space-time which are characteristic for a
gravitational experiment or observation. The result of flat modelling
is in fact a kind of a protocol for a this observation.

Particularly, gravitational effects will manifest themselves as
`discrepancies' in the flat models of closed curves, or, in other
words, as `discrepancies' in observations as they are seen by a
`naive' local observer. For example, closed curves in the real
(curved) space-time may be presented by non-closed flat models.
Another kind of discrepancy which may arise is a Lorentz
transformation of local frame after transporting it along the closed
curve. In both cases the discrepancies in local (flat) presentation of
observations have to be interpreted as evidences of non-zero curvature
of the real space-time.

As examples illustrating the method we shall consider the classical
effect of the deviation of light by Sun and the effect of
gravitational lensing. It will be shown how these effects may be
expressed in terms of flat models of the corresponding closed curves
and compensating Lorentz transformations of local frames.

\section{Flat modelling of a curved space}

It is known that there is no natural mapping of a curved space onto
the flat space (of the same dimension). However, a natural mapping
exists of any curve in a curved space onto a curve in the flat space
of the same dimension. In \cite{FlatModel72} and further publications
of the author \cite{TMF74,PathGr83,PathGr88} this procedure was
exploited to define `flat modelling' of a curved space-time (actually
of arbitrary curves in it). We shall apply this procedure for analysis
of gravitational observations and gravitational effects.

\subsection{Idea of flat modelling and Equivalence Principle}

Consider all curves in a curved space-time $\X$ beginning at the same
point $x_0\in\X$. `Flat models' of these curves are constructed in the
tangent space to the point $x_0$. The tangent space has the structure
of Minkowski space, thus the curve in the curved space-time has a
curve in Minkowski space-time as its model (call it `flat model').
Later on (\Sect{SectMathModel}) we shall describe the process of flat
modelling in precise terms, but now let us give a simple and obvious
idea of this process.

The (tangent) Minkowski space-time will be naturally interpreted as a
geometric image of the whole space-time as it is imagined by a `naive
observer' located at point $x_0$. We thus assume that the `naive
observer' thinks of the real space-time as a flat (Minkowski) one.

Having Minkowski space as a flat model of the space-time, the naive
observer will try to describe gravitational experiments or
observations in terms of this model. Particularly, he will attempt to
present the characteristic curves arising in the course of the
observation as curves in Minkowski space.

Evidently the local observer can do this. For example at a certain
time moment he issues a photon (actually a number of photons) in a
certain direction. Then he constructs the flat model of the world line
of this photon as a light-like straight line in Minkowski space $\M$
starting in the point $\xi_0\in\M$ which is identified with the
location $x_0$ of the observer in the real space-time and pointing to
the corresponding direction. The world line of the observer himself
will be drawn in $\M$ as the straight line starting in $\xi_0$ and
being a trajectory of a body at rest.

Let us assume that at a later time (corresponding, say, to the point
$\xi_1$ at the flat model of his trajectory) the observer sees a
photon (actually a number of photons) coming from a certain direction
(determined by the direction of the telescope's axis). Then he can
construct the flat model of the world line of the absorbed photon.
This will be a light-like straight line coming from the corresponding
direction of the Minkowski space into the point $\xi_1$.

\begin{figure}[ht]
\let\picnaturalsize=N
\def\picsize{2.0in}
\def\picfilename{tangent.eps}
\ifx\nopictures Y\else{\ifx\epsfloaded Y\else\input epsf \fi
\let\epsfloaded=Y
\centerline{\ifx\picnaturalsize N\epsfxsize \picsize\fi \epsfbox{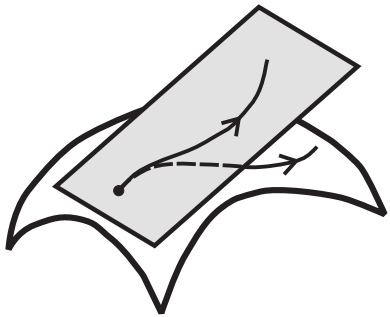}}}\fi
\caption{A curve in a curved space-time and its model in the (tangent)
Minkowski space.}
\label{FigFlatModel}
\end{figure}

Analogously the observer may construct flat models of all curves
crossing the point $x_0$ or any other point of his world line. If he
can by some way or another obtain the information about lengths along
any of these lines and directions of these lines in each point, then
he will naturally construct the flat models of these lines.
(\Fig{FigFlatModel}).
The process of flat modelling is similar to creating a topographical
survey.

In terms of flat modelling, Equivalence Principle may be formulated in
a simple and natural way \cite{FlatModel72,QuEquivPrinc}: the flat
model of the world line of any point-like body acted on by no but
gravitational forces is a time-like straight line in Minkowski space.

\subsection{Gravitational Effects}

If an observer describes his observations by flat models of the
corresponding curves, then gravitational effects manifest themselves
as discrepancies in the resulting picture: the curves which are in
reality closed may be modelled by non-closed curves, and the motion
along these may be accompanied by a certain Lorentz transformation of
the local frame.

\begin{figure}[ht]
\let\picnaturalsize=N
\def\picsize{2in}
\ifx\nopictures Y\else{\ifx\epsfloaded Y\else\input epsf \fi
\let\epsfloaded=Y
{\hspace*{\fill}
 \parbox{3.2in}{\ifx\picnaturalsize N\epsfxsize \picsize\fi
\epsfbox{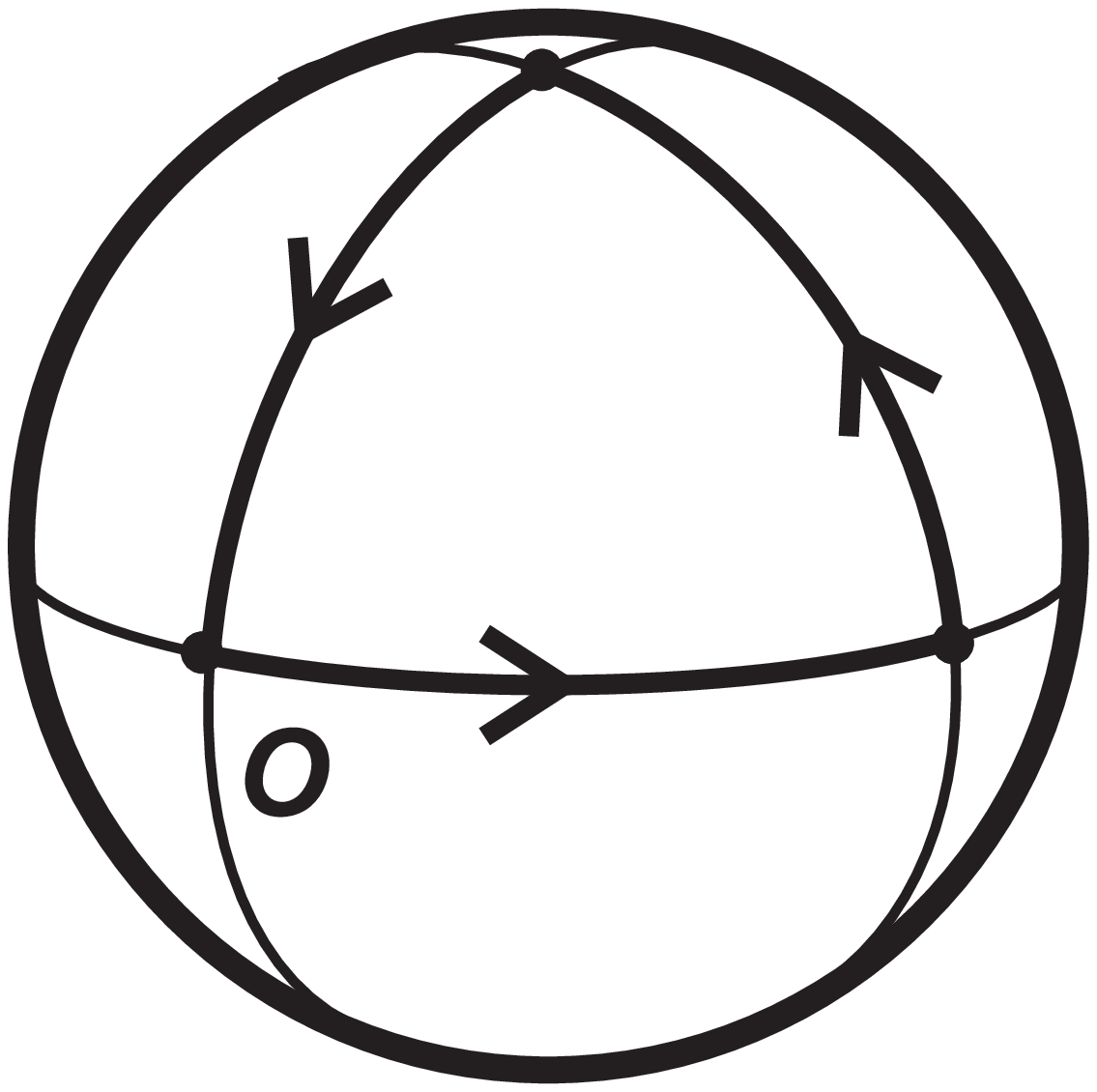}}\hfill
 \parbox{3.2in}{\ifx\picnaturalsize N\epsfxsize \picsize\fi
\epsfbox{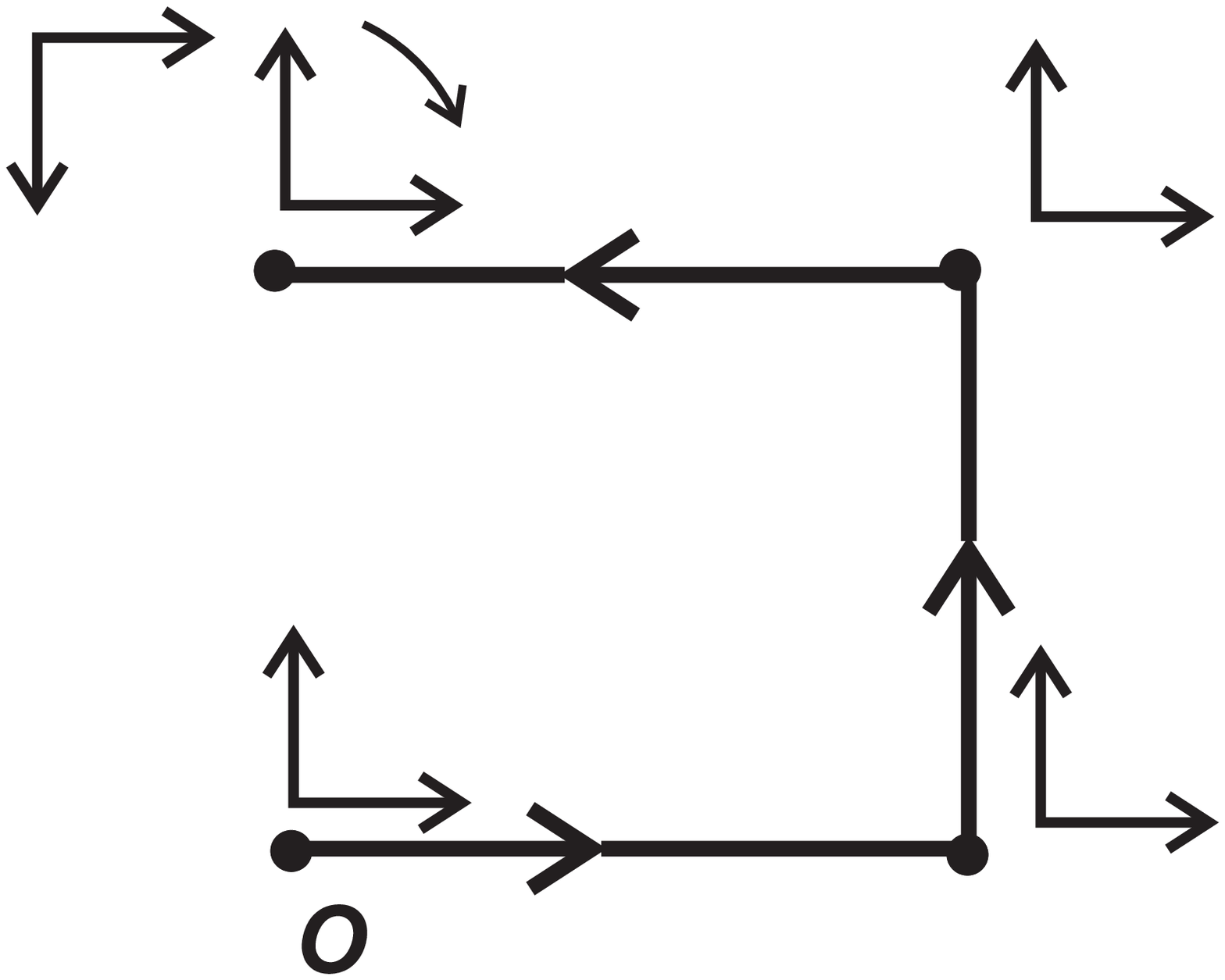}}\hspace*{\fill}}}\fi
\par
\caption{The path on a globe and its flat model.}
\label{Fig-sphere}
\end{figure}

This may be illustrated by the flat model of a curve consisting of
pieces of geodesics on a sphere (\Fig{Fig-sphere}).
Imagine that the sphere is a globe. Let us start from a point on the
equator, go to East along the quarter of the equator, then turn to the
left (to North) and go along the quarter of the meridian (coming to
North pole), then again turn to the left (in the direction of South)
and go along the quarter of the meridian. As a result we return to the
initial point at the equator passing along the closed curve on the
globe.

However, the flat model of this curve will be non-closed (right diagram
of \Fig{Fig-sphere}). It consists of three sides of a square. Besides,
the local frame, after being transported along the globe in a natural
way (this corresponds to the procedure of parallel transport), will
not coincide with the initial local frame. To achieve coincidence, the
transported local frame has to be rotated by the angle $-\pi/2$. Thus,
the flat model of a closed curve is non-closed and needs a
compensating rotation of a local frame. These `discrepancies' are
consequences of non-zero curvature of the sphere. We may consider them
to be manifestations of a `two-dimensional gravitational effect'.

If the flat models and the compensating transformations of local
frames are known for all closed paths with the given initial point,
the geometry may be completely reconstructed (see below
\Sect{SectGeomReconstr}). Real observations cannot of course give this
complete information. Instead, the flat models and compensating
transformations may be known only for restricted number of closed
curves. This determines a class of geometries which lead to the
effect. The complete Information about this class is contained in the
description of those flat models (together with the corresponding
compensating Lorentz transformations) which are obtained from the
observations.

For example, if only a single flat model (with the accompanying
compensating rotation), namely one of the right diagram of
\Fig{Fig-sphere} is known, then the class of geometries which are
compatible with this information consists of deformed spheres which
however coincide with the ideal sphere in the narrow stripe along the
path of the left diagram of \Fig{Fig-sphere}. This may be illustrated
in the following way. Let us cut out, from the right diagram of
\Fig{Fig-sphere}, a narrow stripe of paper along which the flat model
of our curve is located. Let us glue up the end of this stripe with
its beginning after rotating the end by right angle. Then the geometry
corresponding to the observation contains this stripe (glued up in
this way) but is otherwise arbitrary.

\section{Mathematical formalism of flat modelling}\label{SectMath}

Here we shall shortly outline the rigorous mathematical formalism
which is necessary for the construction and analysis of flat models. A
more detailed exposition of this may be found in
\cite{TMF74,PathGr83,PathGr88,QuEquivPrinc}.

\subsection{Flat modelling of curves}\label{SectMathModel}

There is no natural point-to-point mapping of a curved space-time onto
the Minkowski space. However, a {\em natural correspondence exists
between the curves in the Minkowski space and the curves in the curved
space-time} provided the starting point of these curves and local
reference frame in this point are fixed. We shall give here the main
definitions assuming that the (pseudo-Riemannian) geometry is given.

Let $T_x$ be a tangent space in the point $x$ of the curved space-time
$\X$. Then a local frame $b$ in $x$ consists of four tangent vectors,
one time-like and three space-like, $b_{\alpha}\in T_x$,
$\alpha=0,1,2,3$. Each of these vectors $b_{\alpha}$ may be presented
by its components (in any given coordinates) $b_{\alpha}^{\mu}$,
$\mu=0,1,2,3$. The set $\B$ of all local frames (in all points
$x\in\X$) form a fiber bundle over $\X$ as a base. The structure group
of $\B$ is $GL(4)$. The coordinates of points and components of
vectors of local frames $(x^{\mu}, b^{\lambda}_{\beta})$ may serve as
coordinates in $\B$.

A local frame $n\in\B$ is orthonormal if its components are
orthonormalized,
$g_{\mu\nu}\, n_{\alpha}^{\mu}n_{\beta}^{\nu}=\eta_{\alpha\beta}$,
with $g_{\mu\nu}$ being metric in $\X$ and $\eta_{\alpha\beta}$
Minkowski tensor. The set $\N\subset\B$ of all orthonormal local
frames (in all points) is a fiber bundle over $\X$ with the Lorentz
group $\Lambda=SO(1,3)$ as a structure group,
$(n\lambda)_\al=n_{\beta}{\lambda^{\beta}}_\al$. Actually we need only
orthonormal local frames, but the definitions will be simpler if all
frames are introduced at intermediate steps (because it is simpler to
introduce coordinates in $\B$ than in $\N$).

Let us introduce now the so-called {\em basis vector fields} in the
fiber bundle $\B$ of local frames:
\be
B_{\alpha}
=b_{\alpha}^{\mu}\partderiv{}{x^{\mu}}
+b_{\alpha}^{\mu}b_{\beta}^{\nu}\Gamma_{\mu\nu}^{\lambda}(x)
  \partderiv{}{b_{\beta}^{\lambda}}.
\label{Ba}\ee
Here $\Gamma_{\mu\nu}^{\lambda}(x)$ are coefficients of the connection
which for simplicity (but not necessarily) may be taken to coincide
with Christoffel symbols. The vector fields $B_{\alpha}$ are {\em
horizontal} in the fiber bundle $\B$ (in respect to the given
connection) and their restrictions on the fiber bundle $\N$ of
orthonormal local frames are horizontal in $\N$.\footnote{Actually we
need only these restrictions, but it is easier to introduce the field
in $\B$ first.}

Let a curve $[\xi]$ in the Minkowski space $\M$ be given. The
horizontal vector fields enable one to define, as an {\em ordered
exponential} of an integral along this curve, the following operator
acting in the space of functions on $\N$:
\be
V[\xi]=Pe^{\int B_{\alpha}d\xi^{\alpha}}
= \lim_{N\to\infty}e^{B_{\alpha}\Delta\xi_N^{\alpha}} \dots
e^{B_{\alpha}\Delta\xi_1^{\alpha}}.
\label{V}\ee
In terms of these operators, the mapping $n\rightarrow n[\xi]$
may be defined as follows:
\be
(V[\xi]\Psi)(n)=\Psi(n[\xi]).
\label{Vpsi}\ee
This mapping associate the end of the curve $[n]$ with its initial
point. Applying the same procedure to an initial part of the curve
$[\xi]$ (instead of the complete curve), we may obtain also an
arbitrary intermediate point of the curve $[n]$. The projection of
this curve onto the space-time $\X$ (the base of the fiber bundle
$\N$) gives the curve $[x]$ for which $[\xi]$ is a flat model.

This definition may be formulated also in terms of differential
equations. The curve $[x]$ in the curved space $\X$, the corresponding
curve $[n]$ in the fiber bundle $\N$ and the flat model $[\xi]$ of the
curve $[x]$ (i.e. the corresponding curve in Minkowski space $\M$) are
connected by the following differential equations:
\ba
\dot x^{\mu}(\tau)&=&\dot{\xi}^{\alpha}(\tau)n^{\mu}_{\alpha}(\tau),
\nonumber\\
\dot n^{\lambda}_{\beta}(\tau)
  &=&-\dot x^{\mu}(\tau) n^{\nu}_{\beta}(\tau)
  \Gamma^{\lambda}_{\mu\nu}(x(\tau)).
\label{flat-model-dif-eq}\ea
All local frames $n(\tau)$ belonging to the curve $[n]$ are
automatically orthonormal provided that the local frame in the initial
point is chosen to be orthonormal.

\subsection{Path Group, Holonomy Subgroup and reconstruction of
the space-time geometry}\label{SectGeomReconstr}

Geometry of a (curved) space-time $\X$ can be reconstructed if all
elements of the so-called Holonomy Subgroup of the Generalized
Poincar\'e Group are known. In the simplified language used in this
paper this means that the flat models and compensating Lorentz
transformations are known for all closed curves beginning in an
(arbitrary) fixed point $x_0$. We shall present here the general
scheme of the definitions, the details may be found in
\cite{PathGr83,PathGr88}.

Generalized Poincar\'e Group $Q$ is defined as a semidirect product of
the Lorentz group $\Lambda$ and the Path Group $P$, the latter being a
generalization of the translation group, consisting of classes of
curves in Minkowski space (flat models in our context). A generic
element $q$ of $Q$ may be presented as a product, $q = p\lambda$, of a
path $p\in P$ and a Lorentz transformation $\lambda\in \Lambda$.
Products of any elements of $Q$ are defined unambiguously if the
following relation is accepted:
$$
\lambda [\xi] \lambda^{-1} = [\lambda\xi].
$$
The action of paths $p\in P$ on the fiber bundle $\N$ of orthonormal
local frames is defined above (\Sect{SectMathModel}), and Lorentz
transformations act on $\N$ as a structure group. Therefore, the
action of any element of $Q$ on $\N$ is defined naturally as $q: \; n
\rightarrow nq = (n p) \lambda$.

Let us choose an arbitrary orthonormal local frame $n_0\in\N$ which
will play the role of `origin'. Define the Holonomy Subgroup $H\subset
Q$ as the set of elements conserving $n_0$:\footnote{The same
definition but with another choice of $n_0$ would give an isomorphic
Holonomy Subgroup.}
$$
H=\{ h\in Q| n_0 \, h = n_0 \}.
$$
Each element $h\in H$ has the form $h=p\lambda$ where $p\in P$ is a
flat model of some closed curve in the curved space-time $\X$ and
$\lambda\in\Lambda$ a corresponding compensating Lorentz
transformation.

If one knows the whole subgroup $H$ (i.e. flat models and compensating
Lorentz transformations for all closed curves in $\X$ with the given
initial point), then the geometry of $\X$ can be completely
reconstructed. The procedure of reconstruction consists of the
following steps.

\bi
\item The fiber bundle of orthonormal local frames, $\N = H\verb+\+Q$,
is a quotient space of the group $Q$ in respect to the subgroup $H$ so
that a local frame is a coset, $n=Hq=\{hq|h\in H\}$.

\item The space-time $\X$ is a quotient space, $\X = \N/\Lambda$, of
the fiber bundle $\N$ by its structure group $\Lambda$ so that a point
of this space is a set (fiber) of local frames,
$x=n\Lambda=\{n\lambda|\lambda\in\Lambda\}$. Therefore, the mapping $\pi:
n\rightarrow x=n\Lambda$ is the canonical projection of the fiber bundle
$\N$ onto its base $\X$.

\item In order to define the metric on the space $\X$, consider an
arbitrary (orthonormal) local frame $n\in\N$ and the neighboring local
frames $n \, p(a)$ which are obtained from $n$ by action of short
straight paths $p(a)\in P$, coinciding with the four-vectors $a$. Then
the interval between the point $x=\pi(n)$ and $x+dx=\pi(n \, p(da))$
is equal to $ds^2=\eta_{\alpha\beta}da^{\alpha}da^{\beta}$.
\ei

\section{Examples}

Consider two examples of gravitational effects analyzed with the
help of paths (flat models of curves). In each case we shall consider
closed curves characteristic for the observation and construct their
flat models (elements of holonomy). The first example is the
well known experiment on light bending near Sun. The second example is
gravitational lensing.

\subsection{Light bending in terms of paths}

\begin{figure}[ht]
\let\picnaturalsize=N
\def\picsize{2in}
\ifx\nopictures Y\else{\ifx\epsfloaded Y\else\input epsf \fi
\let\epsfloaded=Y
{\hspace*{\fill}
 \parbox{3.2in}{\ifx\picnaturalsize N\epsfxsize \picsize\fi
\epsfbox{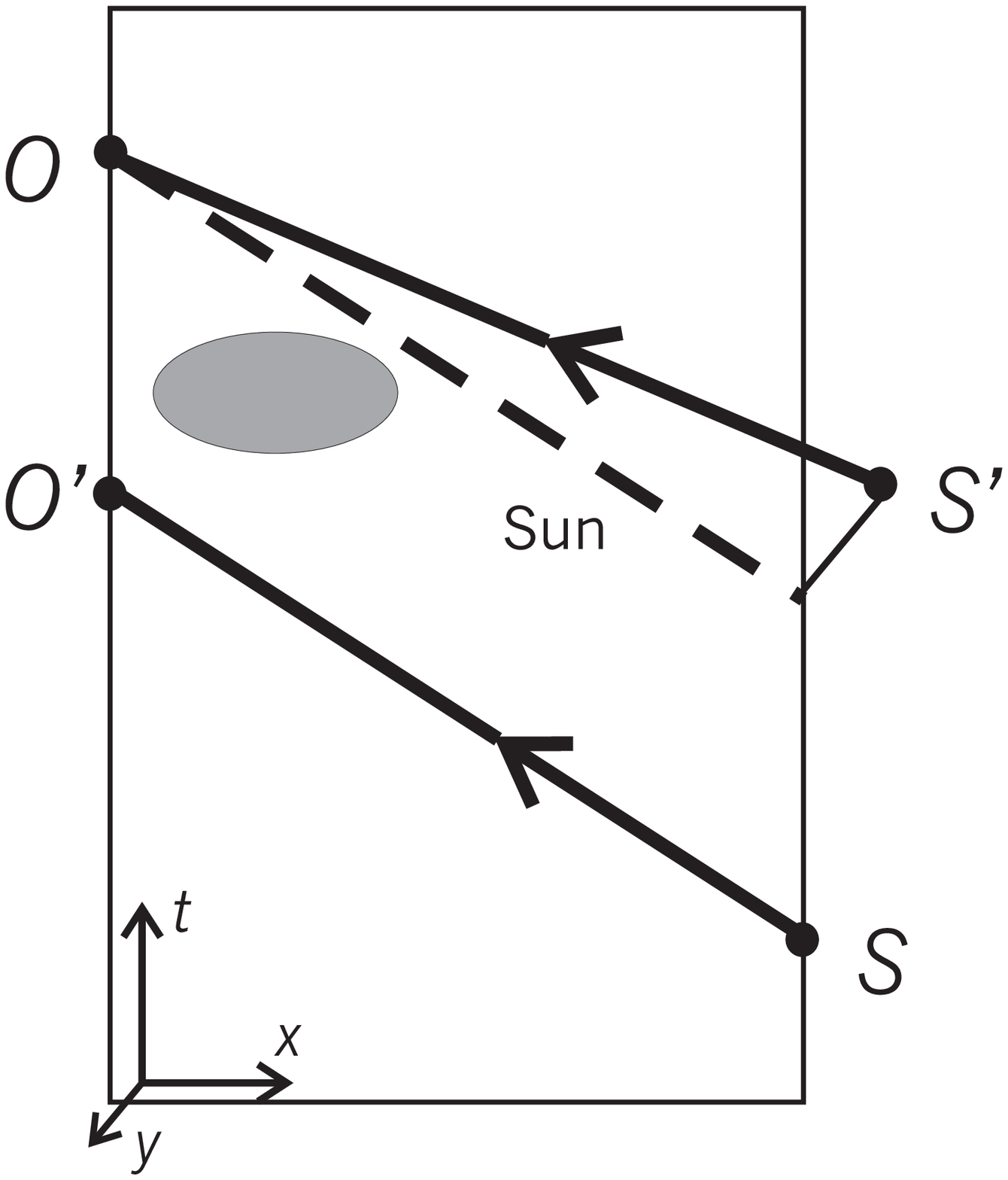}}\hfill
 \parbox{3.2in}{\ifx\picnaturalsize N\epsfxsize \picsize\fi
\epsfbox{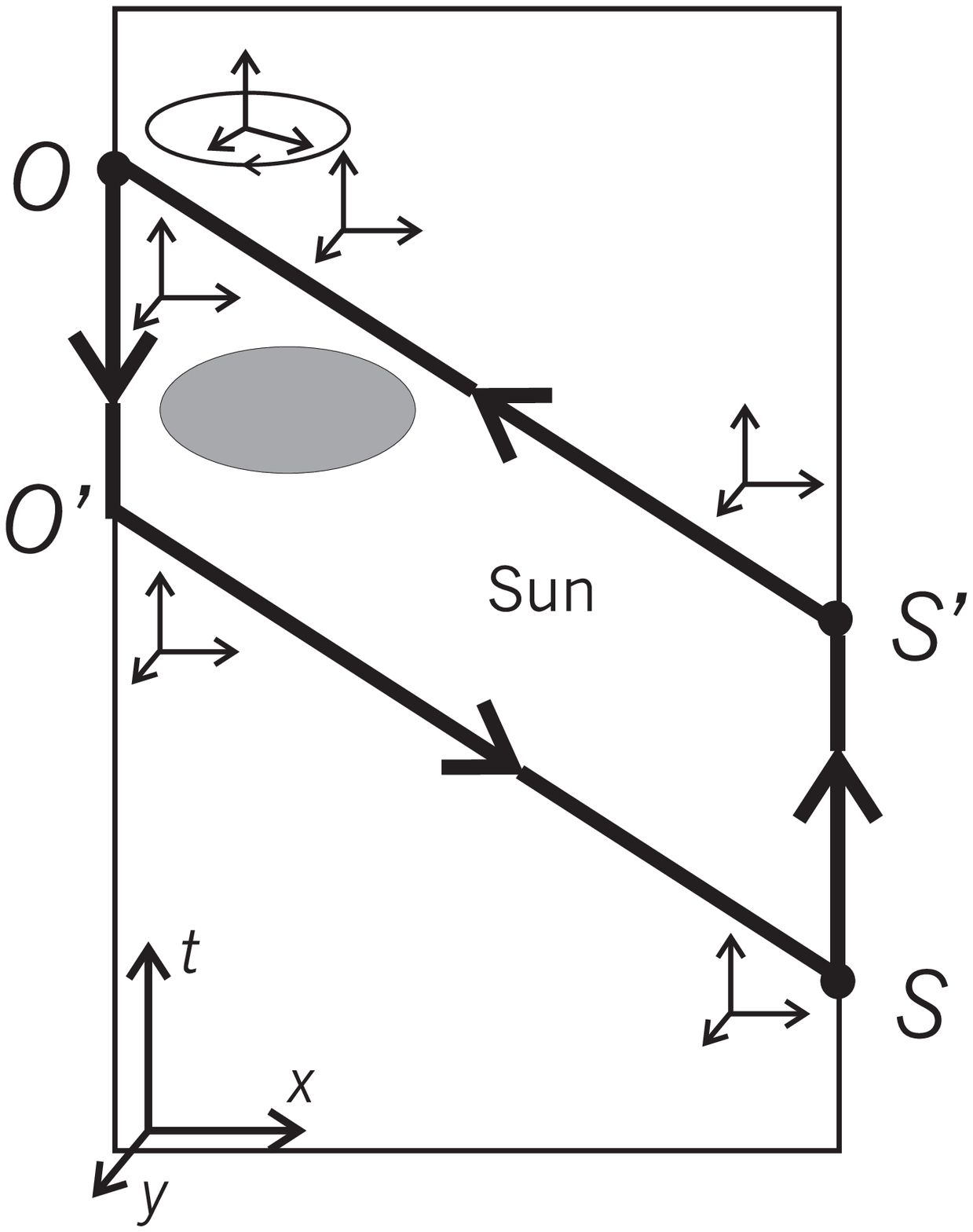}}\hspace*{\fill}}}\fi
\par 
\hspace*{\fill}$a$\hfill $b$\hspace*{\fill}\\
\caption{Light bending by Sun and the corresponding element
of holonomy.}
\label{Fig-ah241730}
\end{figure}

The effect of light bending by Sun is easily analyzed by the
conventional method. We shall speak about this effect only to
illustrate our method. In the observation of light bending the
direction of the light ray issued by a star turns out to be different
in two cases: when this ray passes near Sun and when it passes far
from Sun. It is shown at \Fig{Fig-ah241730}a
how the observer construct flat models of the world lines of the two
photons seen in the two points of his world line (flying
correspondingly near Sun and far from it). For the analysis of the
gravitational effect a closed curve is necessary. It may be formed by
the world lines of the two photons together with the world lines of
the observer and the star.

In order to obtain a flat model of this curve, one needs to assume
that the only gravitational effect is caused by Sun deflecting the
photon. If this hypothesis is accepted, then the flat model of the
closed curve may be reconstructed as it is shown in
\Fig{Fig-ah241730}b. The flat model of the closed curve (i.e. the path
$p$ in the expression $h=p\lambda$ of the element of the Holonomy
Subgroup) is also a closed curve consisting of two light-like lines
and two world lines of the bodies at rest.\footnote{Actually the path
$p$ is non-closed, but this effect is very small and may be neglected
because the distance from the observer to the star is much longer than
to Sun.}

The observed effect is a consequence of the fact that the local frame
is rotated during its parallel transport along the closed curve. To
make the local frame identical to the initial one, it has to be
rotated after the parallel transport. This is a compensating Lorentz
transformation $\lambda$ in the expression $h=p\lambda$ of the element
of holonomy.

\subsection{Gravitational lensing}

\begin{figure}[ht]
\let\picnaturalsize=N
\def\picsize{2in}
\ifx\nopictures Y\else{\ifx\epsfloaded Y\else\input epsf \fi
\let\epsfloaded=Y
{\hspace*{\fill}
 \parbox{2in}{\ifx\picnaturalsize N\epsfxsize \picsize\fi
\epsfbox{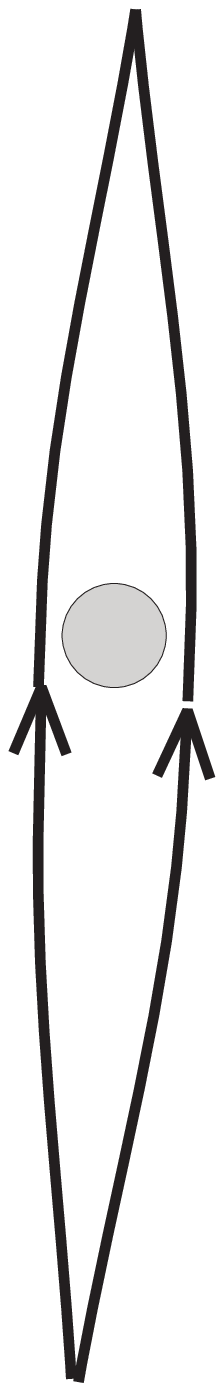}}\hfill
 \parbox{2in}{\ifx\picnaturalsize N\epsfxsize \picsize\fi
\epsfbox{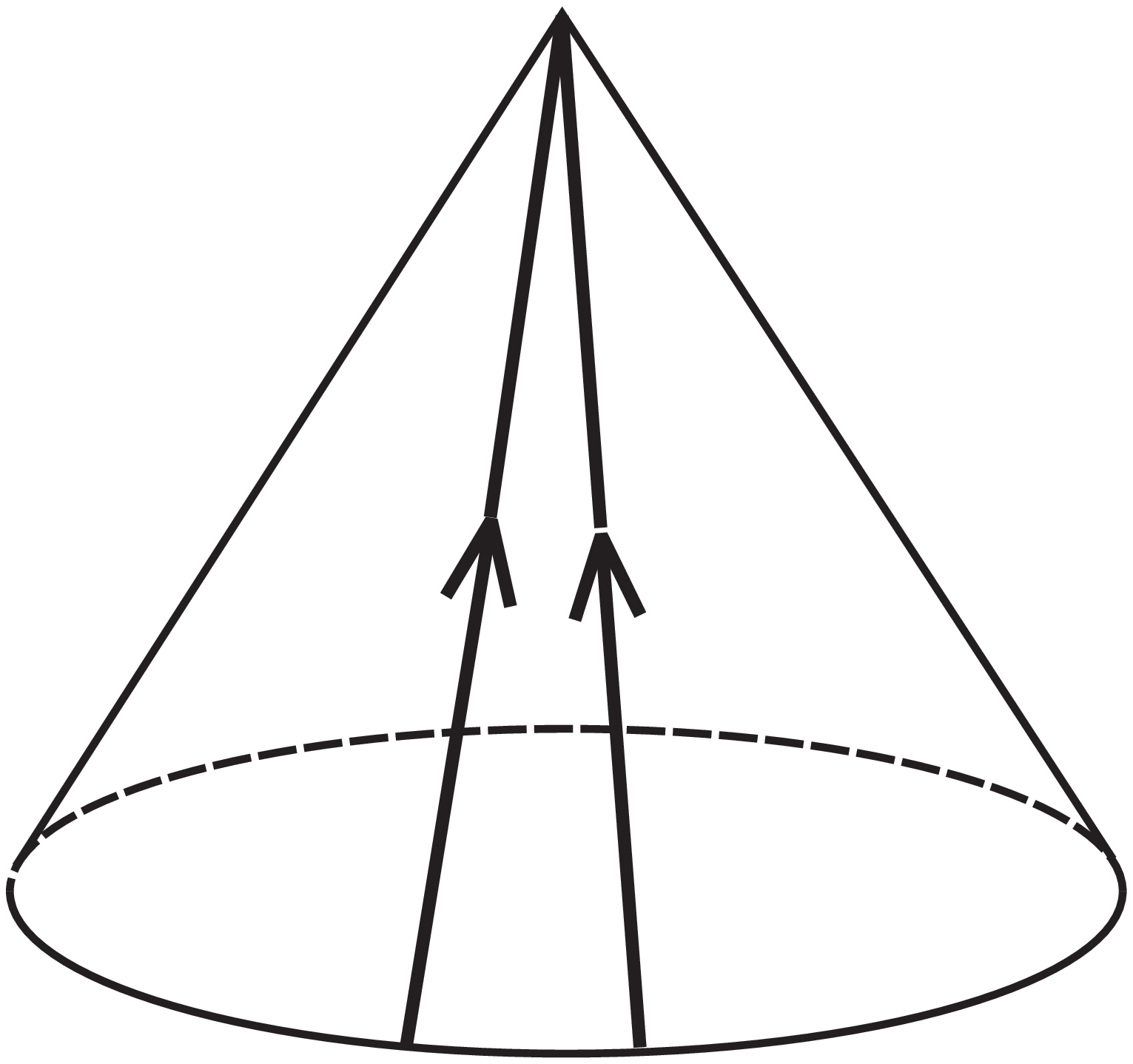}}\hfill
 \parbox{2in}{\ifx\picnaturalsize N\epsfxsize \picsize\fi
\epsfbox{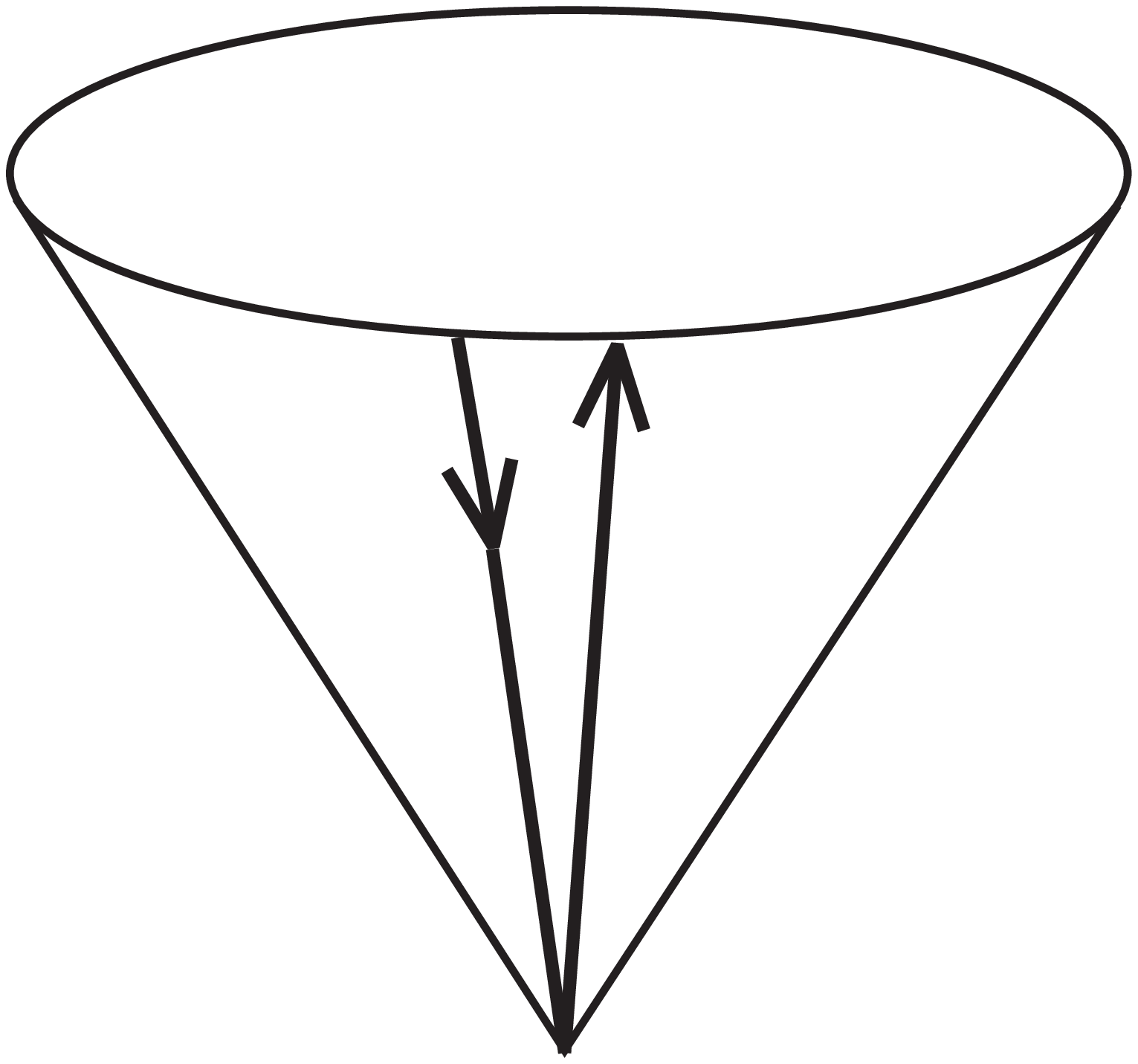}}\hspace*{\fill}}}\fi  
\par
\hspace*{\fill}$a$\hfill $b$\hfill $c$\hspace*{\fill}\\
\caption{Gravitational lensing}
\label{FigLense}
\end{figure}

In case of gravitational lensing the observer sees light issued by
some object (a star) as if coming from two (or more) different
directions (\Fig{FigLense}a).
Trajectories of photons issued by the star are curved because of
being attracted by a heavy object located between the star and the
observer. As a result, the observer sees two separate light spots
in different directions. The flat model of the world lines of the two
photons passing the attracting object from different sides will be as
in \Fig{FigLense}b: two light-like lines belonging to the past light
cone of the observer.

However, with the help of the analysis of the radiation the observer
knows that the two photons are in fact issued simultaneously from the
same point. The result of this analysis is that the passage along one
of these lines to the past leads to the event of radiation of the
photon by the star. Passing from this event along the world line of
the second photon to the future returns to the initial event of the
perception of both photons by the observer.

The line consisting of two world lines of two photons is therefore
closed. Its flat model is shown in \Fig{FigLense}c and presents the
path $p$ in some element of holonomy $h=p\lambda$. The above arguments
do not give the value of the compensating Lorentz transformation. One
may assume that $\lambda=1$ so that the local frame parallelly
transported along the given closed curve will be identical with the
initial one. In principle astrophysical data may give the information
about $\lambda$ too, for example if the source issues polarized
photons.

\section{Conclusion}

We shortly discussed here the way of objective presentation of
observations of gravitational effects. The analysis was based on the
mapping of curves in a curved space-time onto curves (paths) in
Minkowski space. The most adequate development of such a formalism is
presented in terms of Path Group, Generalized Poincar\'e group and
Holonomy Subgroup \cite{PathGr83,PathGr88}. In correct terms of this
formalism the observational data provide some elements of the Holonomy
Subgroup, but we preferred to use also a simple image of `flat models' of
closed curves in a curved space-time (but a more rigorous terminology
in \Sect{SectMath}).

It is important that the real observations give only some elements of
holonomy (flat models of some closed curves) while the precise
reconstruction of the geometry is possible only from the knowledge of
the whole Holonomy Subgroup (flat models of all closed curves).
Knowing a restricted set of elements of holonomy (as they are obtained
from the observations), one may point out a class of geometries having
these elements in its Holonomy Subgroup. The presentation of this
class by paths and compensating Lorentz transformations (as in the
examples considered here) is adequate because it is not based on any a
priori choice of the class of geometries.

Let us remark that the formalism of Path Group and Generalized
Poincar\'e Group has also other applications, among them the
application to Quantum Equivalence Principle \cite{QuEquivPrinc}. This
gives an evidence that the formalism is natural in gravity and
particularly in quantum gravity.

\quad

\centerline{\bf ACKNOWLEDGEMENT}

The author acknowledges fruitful discussions with H.von~Borzeszkowsky.
The work was supported in part by the Deutsche Forschungsgemeinschaft.

\end{document}